\documentclass[twocolumn,secnumarabic,amssymb, nobibnotes, aps, prd]{revtex4}

\usepackage{amsmath}
\usepackage{euscript}
\usepackage{graphicx}
\usepackage[english]{babel}

\begin{document}

\title{Using metallic photonic crystals as visible light sources}

\author{Sergei Belousov}
\author{Maria Bogdanova}
\author{Alexei Deinega}
\author{Sergey Eiderman}
\author{Ilya Valuev}
\author{Yurii Lozovik}
\author{Ilya Polischuk}
\author{Boris Potapkin}
\affiliation{Kintech Lab Ltd., 1 Kurchatov Sq., Moscow, Russia 123182}

\author{Badri Ramamurthi}
\author{Tao Deng}
\author{Vikas Midha}
\affiliation{GE Global Research Center, 1 Research Circle, Niskayuna, NY, USA 12309}

\date{\today}

\begin{abstract}
In this paper we study numerically and experimentally the possibility of using metallic photonic crystals (PCs) of different geometries (log-piles, direct and inverse opals) as visible light sources.
It is found that by tuning geometrical parameters of a direct opal PC one can achieve substantial reduction of the emissivity in the infrared along with its increase in the visible.
We take into account disorder of the PC elements in their sizes and positions, and get quantitative agreement between the numerical and experimental results. We analyze the influence of known temperature-resistant refractory host materials necessary for fixing the PC elements, and find that PC effects become completely destroyed at high temperatures due to the host absorption. Therefore, creating PC-based visible light sources requires that low-absorbing refractory materials for embedding medium be found.
\end{abstract}


\maketitle

\section{Introduction}

Photonic crystals (PCs) are artificial structures characterized by periodical variation of the dielectric function in space. Their optical properties differ from the optical properties of continuous media: electromagnetic waves in the PC have a band structure similar to that of electrons in a solid crystal~\cite{PCB}. In particular, the ability of a PC to modify the thermal emission has drawn much attention recently~\cite{Cornelius,FlorescuDOS,FlemingNature,Lin1538,LinAPL1,LinAPL2,LiJAP,OptExp2010,Wan,LKKRem,Han,Rephaeli,SaiGrating,Chan,Celanovic,Araghchini}.
Generally, the thermal emission from a medium at a certain wavelength is determined by an interplay between the photonic density of states
and the energy transport velocity~\cite{Cornelius,FlorescuDOS}.
Thus the thermal emission from a PC exhibits strong wavelength dependence, for example,
it may be suppressed for wavelengths corresponding to band gaps, where
the photonic density of states is zero, or stop bands where it is zero for certain wave directions.
By changing geometric parameters of a PC, one can tune the high or low emissivity wavelength ranges to the desired parts of the spectrum.
This effect has been studied theoretically and experimentally for different PC geometries, such as log-piles~\cite{FlemingNature,Lin1538,LinAPL1,LinAPL2,LiJAP,OptExp2010}, opals~\cite{Wan,LKKRem}, inverse opals~\cite{Han}, pyramids~\cite{Rephaeli}, inverted pyramids~\cite{SaiGrating}, arrays of holes~\cite{Chan,Celanovic,Araghchini}, etc.

One of the most important applications for PC-modified thermal emission is
thermophotovoltaics~\cite{FlemingNature}.
There the spectrum of an emitter should be adjusted to the photovoltaic cell sensitive wavelength range in the infrared (IR)~\cite{TPV}.
Additionally to PC bandstructure effects, possible further ways to enhance efficiency of thermophotovoltaic systems can be based on plasmonic near-field emitters~\cite{Ilic,Pendry}, negative index metamaterials~\cite{Maks,Wang}, intersubband transitions in multiple quantum wells tailored with photonic-crystal resonant effects~\cite{Zoysa}.
Photovoltaic cell by itself can be designed as a PC with improved solar power conversion efficiency which is achieved due to enhanced absorption~\cite{Gui} and applying curved pn-junction~\cite{GuiTr}.


Recently it was proposed to use PCs as a source of visible light~\cite{FlemingNature,Lifcc,SJ2,Narayana}. Suppression of the IR emission of such sources may increase their efficiency in comparison with conventional incandescent lamps which emit mostly (about 95\%) in the IR.
Some candidate PC geometries have been studied theoretically so far.
It was shown that it is possible to achieve high emissivity in the visible range using one-dimensional tungsten photonic crystal, however, emission in the near IR appears to be high as well~\cite{Narayana}.
Emission in the IR could be suppressed using iridium inverted square-spiral photonic crystal~\cite{SJ2}, but fabrication and use of such a structure as a light source is impractical due to high costs of iridium.

In this work we focus on practically achievable PC-based visible light emitters, namely the three- and two-dimensional PC structures with tungsten emitting elements. Tungsten is widely used as a traditional light emitting material for incandescent  lamps due to its high melting temperature ($3695K$) and high selectivity of emission in the visible range. We analyze a number of possible three-dimensional tungsten PC geometries (log-piles, direct and inverse opals), perform numerical simulations, and support the obtained results experimentally.

The rest of the paper is organized as follows.
In section~II, we describe the method for the emissivity calculations and briefly discuss its accuracy.
In section~III, we show that log-piles have a poor emission selectivity in the visible and cannot be used for designing efficient light sources.
In section~IV, we find that by choosing appropriate geometrical parameters of a direct opal PC, one can achieve low emissivity in the IR and high emissivity in the visible.
In section~V, we study the effect of position and size disorder of the PC elements and present comparison with experimental results.
In section~VI, we take optical properties of the refractory host medium necessary for fixing the direct opal PC elements into account.
In the final section~VII, we summarize our results.

\section{PC emissivity calculation method}


According to Kirchhoff's law of thermal radiation, the emissivity of a structure is equal to its absorption coefficient in thermodynamic equilibrium~\cite{Cornelius}. This law may be used for obtaining the emissivity in a numerical experiment by calculating absorption
instead of simulating radiation from a structure. Direct numerical simulation of a radiating body is possible by
various techniques, for example, by using stochastic Langevin electrodynamics~\cite{ChanDirect,Luo} or by truncating
the natural modal expansion of electromagnetic fields inside a weakly absorbing PC~\cite{Schuler}. In both cases rigorous checks
of Kirchhoff's law were performed to prove the consistency of the proposed models for radiation in thermodynamic equilibrium.


In the present work we apply the finite-difference time-domain (FDTD) method~\cite{Taflove} to obtain the absorption
coefficient of a structure numerically as a function of wavelength. The simulation software
was especially created for this problem based on our parallel Electromagnetic Template Library~\cite{EMTL}.
The use of FDTD has a number of
advantages for simulating PC-like structures. It is able to handle purely periodic structures with 2D or 3D periodicity as well as
quasi-periodic geometries with large number of randomly disordered unit cells. For large simulations the method
is parallelized via domain decomposition and scales almost linearly with the number of used CPUs~\cite{EMTL}.
Absorbing materials are simulated
by fitting the experimental complex dielectric function by a number of Drude-Lorentz~\cite{Taflove} or modified Lorentz~\cite{eff_si} terms.
There is no restriction on the form of scatterers in FDTD, however a special treatment of fine structures such as thin
material interfaces is sometimes required~\cite{Tensor}.

For numerical experiment we use a conventional FDTD calculation scheme in which a simulated PC structure, defined by the complex dielectric function of the material is placed in the simulation volume. We use the orthogonal
uniform Yee mesh for the whole volume.
A plane wave pulse is generated on a surface outside the structure with the help of the Total Field/Scattered Field technique and propagates through it (Fig.~\ref{log_Sandia}, inset).
Generally, a PC consisting of several layers stacked in the Z-direction is placed inside a computational volume.
The plane wave pulse impinges on the PC at given direction characterized by the incident angles $\theta$ and $\phi$.
We simulate oblique incidence of a plane wave on the PC by the iterative FDTD algorithm~\cite{Oblique}.
The computational volume is terminated from both sides in the Z direction by absorbing boundary conditions - perfectly matched layers (PMLs)~\cite{Taflove}.
They simulate withdrawal of the transmitted and reflected waves to the infinity. Periodic boundary conditions are applied along the X- and Y-borders of the computational volume. A residual numerical reflection from the PMLs is reduced by applying additional back absorbing layers technique~\cite{PML}.
During the numerical experiment, the reflected and transmitted waves are recorded behind and in front of the structure, transformed to the frequency domain and normalized to the incident spectrum. Ultimately, the transmission ${\EuScript T}(\omega,\theta,\phi)$, reflection ${\EuScript R}(\omega,\theta,\phi)$, and absorption ${\EuScript A}(\omega,\theta,\phi)=1-{\EuScript T}(\omega,\theta,\phi)-{\EuScript R}(\omega,\theta,\phi)$ coefficients are calculated. The emissivity is then obtained by applying Kirchhoff's law as ${\EuScript E}(\omega,\theta,\phi)={\EuScript A}(\omega,\theta,\phi)$.

Throughout this paper the following FDTD parameters are used: a) for the case of log-pile PCs, the mesh resolution is 100 steps per the PC unit cell ($\Delta r = 42$nm for the log-pile with the lattice constant $a = 4.2 \rm \mu m$ and $\Delta r = 5$nm~--~for $a = 500$nm, see Section III), time step $c \Delta t = 0.5 \Delta r$, the numerical experiment duration is $N_T = 40000$ iterations; b) for the case of opal PCs (typical PC parameters: lattice constant $a \sim 500$ nm, sphere radius $R \geq 50$nm), $\Delta r=2.5$nm, time step $c \Delta t = 0.5 \Delta r$, the numerical experiment duration is $N_T=20000$ iterations.

FDTD is a well established method for simulating wave propagation. The accuracy of any particular simulation should however be
checked by the analysis of its convergence with respect to the decreasing mesh step. In cases of poor convergence, special FDTD
techniques, for example subpixel smoothing for
interfaces or additional PML layers may be required. Convergence and accuracy of FDTD calculations based on EMTL
 have been demonstrated in our previous papers~\cite{EMTL,Oblique,Tensor,PML}.
In particular, we considered a general case of an obliquely incident plane wave on a metallic opal PC~\cite{Oblique}, and demonstrated an excellent agreement of the FDTD results with the results obtained by the multiple scattering formalism~\cite{LKKR,LKKR2}. Also, the opal PC case requires a special care as the spherical shape of the PC elements should be adequately represented on the rectangular FDTD grid. In our previous work we presented a subpixel smoothing method based on the inverse dielectric tensor approach~\cite{Tensor}. This method has been shown to significantly improve the accuracy of the FDTD simulations for arbitrary shaped metal scatterers and is used in the present work for the opal PCs simulations.


In the present work tungsten is considered as the emitting material for PC light sources.  The data on the tungsten dielectric function $\varepsilon(\omega)$ in Drude-Lorentz form is taken from~\cite{Roberts}. Our numerical results are obtained for the temperature $T = 2400 K$ (if not stated otherwise) which is the typical working temperature of a tungsten filament in incandescent lamps.

Energy flow from a PC layer with the emissivity ${\EuScript E}(\omega,\theta,\phi,T)$ operating at an ambient temperature $T$ within a given frequency range $\omega_1 \le \omega \le \omega_2$ is calculated as
\begin{equation}
J(T)= \int\limits_{\omega_1}^{\omega_2} d\omega \int\limits_0^{2\pi} d\phi \int\limits_0^{\pi/2} {\EuScript E}(\omega,\theta,\phi,T) u(\omega,T) \cos{\theta} \sin{\theta} d\theta, \label{pc_J}
\end{equation}
where $u(\omega,T)$ is the blackbody spectral radiation density
\begin{equation}
u(\omega,T)=\frac{\hbar\omega^3}{4\pi^3c^2}\frac{1}{\exp{(\hbar\omega/kT)}-1}.
\end{equation}
The efficiency of light emission $E$ is obtained as the ratio of the emitted energy flow in the visible $J_{\rm vis}$ to the total energy flow $J_{\rm tot}$.

Note that typical thickness of a tungsten glower~($\sim 0.2 \mbox{mm}$) is much larger than emission wavelength (both in visible and IR).
Therefore to calculate the efficiency of the glower one can simulate it as a semi-infinite tungsten layer. For such a layer, we obtain an efficiency of $E_{\rm W}= 5\%$ and emitted energy flow in the visible $J_{\rm vis,W}=30 \mbox{kWt/m}^2$ (at $T = 2400 K$). We use this values as references characterizing
a standard incandescent lamp.
Our purpose is to find such a tungsten PC geometry which yields the efficiency $E$ several times greater than $E_{\rm W}$, keeping $J_{\rm vis}$ of the same order as $J_{\rm vis,W}$ at the same time, so as to provide sufficient emitted energy in the visible.

\section{Emissive properties of log-piles PC}

Metallic log-pile structures have drawn a special attention due to their possible use in thermophotovoltaics~\cite{FlemingNature} as IR emitters. Particularly, tungsten log-piles providing high emission selectivity in the near IR range were demonstrated~\cite{Lin1538,LinAPL1,LinAPL2,LiJAP}. These structures are characterized by a photonic band gap in the IR range (we refer to it as the IR band gap later on), and a stop band in $\Gamma X$ direction~\cite{FlemingNature,Lin1538,OptExp2010}.

We note at this point, that the terms "band gap" and "stop band" are related to an extended 3D PC
rather than to finite PC samples (composed of a finite number of layers). However, as the thickness of a PC increases, band gaps and stop bands start manifesting themselves as high reflectivity regions. It has been shown~\cite{Lin1538}, that in the case of a metallic log-pile, as much as only four layers are already sufficient for manifestation of the band structure features. Because of this correspondence we will continue using the band structure terminology throughout the current section when discussing the high reflectivity (low transmittance) regions in the spectra of a finite PC.

\begin{figure}
  \vspace{-10pt}
  \centering
  \scriptsize
  \includegraphics[width=1\linewidth]{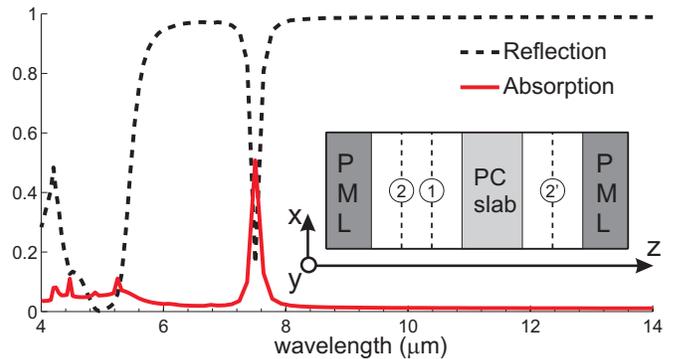}
  \caption{\small Normal incidence reflection and absorption spectra of a log-pile with number of layers $N = 4$, rod spacing $a = 4.2 \mu m$, rod width $w = 1.2 \mu m$, and rod height $h = 2.4 \mu m$. High reflectance region at $\lambda > 8\mu m$ correspond to the IR band gap. Sharp peak appears at the IR band gap edge. Inset: scheme of the FDTD simulation geometry; $1$ -- generating (Total Field / Scattered Field) border; $2$, $2'$ -- detector arrays for reflected and transmitted signals.}
  \label{log_Sandia}
\end{figure}

Let us illustrate the properties of the tungsten log-pile PC operating in the IR range on Fig.~\ref{log_Sandia} by FDTD simulation. We use the same geometrical parameters, as in~\cite{Lin1538}: the lattice constant $a = 4.2 \mu m$, the rod width $w = 1.2 \mu m$, the rod height $h = 2.4 \mu m$ and the number of layers $N = 4$. The dielectric function of tungsten at $T = 298 K$ is used.
The calculated reflectance and absorbance spectra are shown in Fig.~\ref{log_Sandia}.
This PC has two high reflectivity regions, corresponding to the IR band gap and the stop band~\cite{Lin1538}, where absorption and, therefore, thermal emission is suppressed, and exhibits a sharp absorption (emissivity) peak between the IR band gap and the stop band due to the interplay between the low light group velocity and high density of states ~\cite{Cornelius,FlorescuDOS,Lin1538,OptExp2010}. The high emission selectivity in the near IR is a direct consequence of this peak. Basing on the above results, it was suggested~\cite{FlemingNature} that by a proper scaling of the log-pile geometry, all spectral features could be shifted to the visible range, thus opening the possibility of using log-piles as efficient light sources.

\begin{figure}
  \vspace{-10pt}
  \centering
  \scriptsize
  \includegraphics[width=1\linewidth]{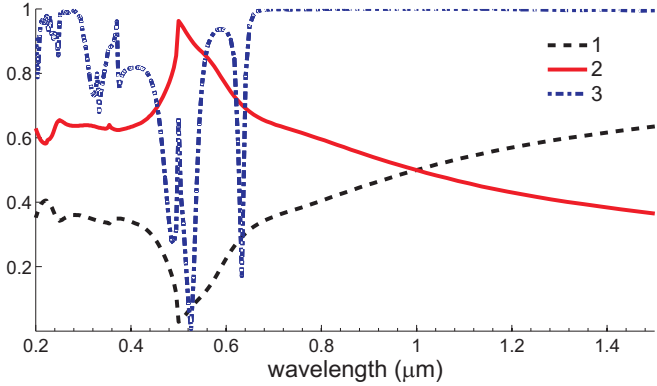}
  \caption{\small Normal incidence reflection (1) and absorption (2) spectra of a tungsten log-pile at $T = 2400 K$ with $N = 4$, $a = 0.5 \mu m$, $w = 0.2 \mu m$, and $h = 0.2 \mu m$. The reflection for the perfect metal PC is shown for reference (3).}
  \label{log_thick}
\end{figure}

However, our findings indicate, that high intrinsic losses of tungsten in the visible range wash out the relevant spectral features of the scaled log-pile geometry. We compare the simulation results for 4 layer perfect metal (no losses) and tungsten log-piles with scaled geometrical parameters ($a = 0.5 \mu m$, $w = 0.2 \mu m$, $h = 0.2 \mu m$) chosen so as to move the IR band gap edge to the visible range. While the behaviour of the reflectance with the wavelength for the perfect metal log-pile is very similar to the one seen in Fig.~\ref{log_Sandia}, the case of the tungsten log-pile is different. Its absorption spectrum has a single broad peak at the $\lambda \sim 0.5 \mu$m and falls gradually at the longer wavelengths, remaining quite high in the near IR (Fig.~\ref{log_thick}). Correspondingly, the reflectance of the log-pile increases in the near IR, indicating the IR band gap formation.
The stop band is completely destroyed while the IR band gap is strongly affected by the high absorption of tungsten, leading to the low emission selectivity in the visible range.
Similar results have been obtained in the recent experimental study of metallic log-piles~\cite{WPopt}.


\begin{figure}
  \vspace{-10pt}
  \centering
  \scriptsize
  \includegraphics[width=1\linewidth]{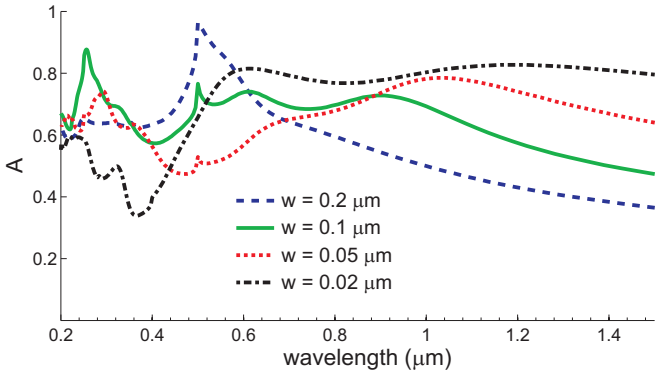}
  \caption{\small Normal incidence absorption spectra of a tungsten log-pile at $T = 2400 K$ with $N = 4$, $a = 0.5 \mu m$, $h = 0.2 \mu m$, and varying $w$.}
  \label{log_thin}
\end{figure}

\begin{figure}
  \vspace{-10pt}
  \centering
  \scriptsize
  \includegraphics[width=1\linewidth]{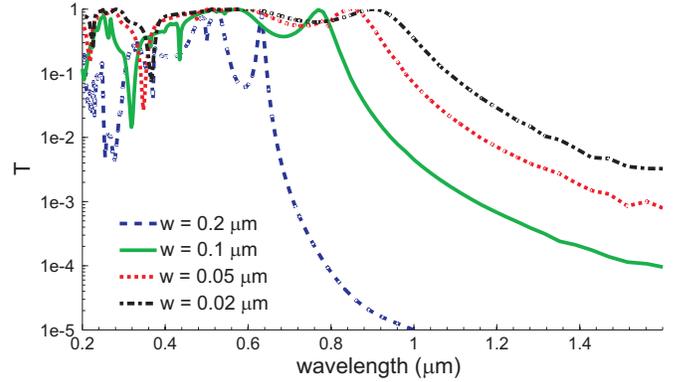}
  \caption{\small Normal incidence transmission spectra of a perfect metal log-pile with $N = 4$, $a = 0.5 \mu m$, $h = 0.2 \mu m$, and varying $w$.}
  \label{log_thin_perfect}
\end{figure}

Naturally, one could think of reducing the tungsten filling fraction in order to compensate for tungsten losses, which could be achieved by, e.g., decreasing the width of the rods. We consider the tungsten log-pile with a varying rod width $w$, keeping both $a$ and $h$ constant. Simulations reveal, however, that, contrary to the intuition, the absorption increases in the near IR with decreasing the log width $w$ (Fig.~\ref{log_thin}). This behaviour can be better understood by considering the transmission spectra of the perfect metal log-piles. In this case, decreasing $w$ with both $a$ and $h$ kept constant, leads to the red shift and weakening of the band gaps (Fig.~\ref{log_thin_perfect}). This can be qualitatively associated with the grating-like geometry of each layer of the log-pile. Such a grating is characterized by a waveguide cutoff wavelength, corresponding to twice the air opening of the grating $\lambda_{c} = 2(a-w)$~\cite{LiJAP}. The transmittance is strongly attenuated for $\lambda>\lambda_{c}$, indicating the band gap formation. Decreasing $w$ leads to the increasing of the air opening along with the filling factor decrease. This results in both the red-shift and the weakening of the gap, allowing more radiation inside the structure due to the higher transmittance. This, in turn, leads to the higher absorption in the near IR range in case of the lossy tungsten log-pile, which is exactly what we have observed in (Fig.~\ref{log_thin}).


In summary, the intrinsic tungsten absorption strongly affects the spectral characteristics of the log-piles, hampering their application as light sources.
\section{Emissive properties of opal PC}

\begin{figure}
  \vspace{-10pt}
  \centering
  \scriptsize
  \includegraphics[width=1\linewidth]{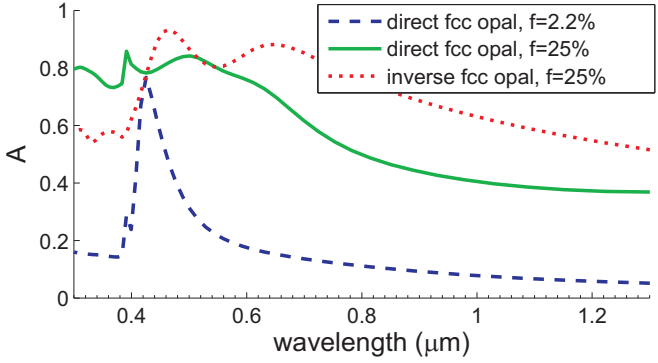}
  \caption{\small The absorption spectra of direct and inverse FCC opals at normal incidence and varying filling fraction $f$. The lattice period $a=0.45$ $\mu m$, number of layers $N=2$, $T=2400K$. The host medium is vacuum.}
  \label{kirch_types}
\end{figure}

We now turn to discussing the emissive properties of opal-like PCs.
We consider a PC slab consisting of $N$ layers of spheres arranged in the FCC lattice, stacked along (111) crystallographic direction.
Each layer of this structure is a 2D triangular lattice with the period equal to $a$ (FCC lattice constant is $\sqrt{2}a$).
There are two possible opal geometries: direct (tungsten spheres in air) and inverse (air spheres in tungsten slab of the finite thickness) opals. Here air should be replaced in practice by a refractory supporting of filling host material, but we start from numerically studying the effect of a tungsten PC itself.   A low tungsten filling fraction can be achieved with the direct opal geometry, since the sphere radius can be made as small as necessary. As we show in the current section, low filling fraction leads to higher efficiency in the visible. As an illustration, we present calculated absorption spectra for direct FCC opals with filling fractions of 2.2\% and 25\% and inverse FCC opal with filling fraction of 25\% (Fig.~\ref{kirch_types}).
The absorption spectrum of the direct opal with $f = 2.2\%$ is characterized by a pronounced peak at 500nm, with amplitude several times greater than the absorption in the IR.
The high absorbance ranges are also present in the spectra of both direct and inverse opals with $f = 25\%$, but they are broadened with respect to the low filling fraction case, and their maximum amplitude exceeds the IR absorption by only a factor of 1.5.
Note that in inverse opals with $f < 25\%$ the necks between the air cavities partially vanish. At high temperatures they are likely to sinter which results in thermal instability of this structure.
This makes inverse opals less favorable for light applications.
In view of the above arguments, we confine with studying the emissivity properties of direct opals with the low filling fraction. Below we study how their emissivity depend on the lattice period $a$, radius $R$ and number of layers $N$.

\begin{figure}
  \vspace{-10pt}
  \centering
  \scriptsize
  \includegraphics[width=1\linewidth]{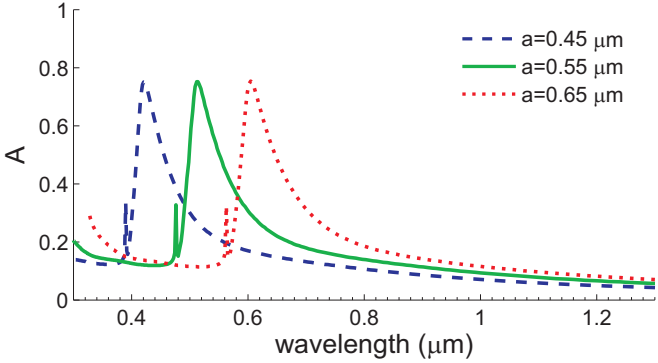}
  \caption{\small The absorption spectra of PC at normal incidence for varying lattice constant $a$, and $R=0.156 a$. $N=2$, $T=2400K$.
  }
  \label{kirch_dif_a}
\end{figure}

In Fig.~\ref{kirch_dif_a} we present the calculated absorption for a two-layered direct opal PC.
while scaling its geometrical parameters.
There is a pronounced peak in the absorption spectra, scaled linearly with $a$ (if $R/a$ is constant).
By adjusting the $a$ value, one can tune the absorption peak position into the visible range.
The nature of this peak is related to the resonant absorption of the electromagnetic waves in the 2D triangular layers constituting the opal PC.
At the normal incidence, the resonant absorption condition is fulfilled for a wavelength $\lambda$  when $2\pi/\lambda$ is close to the magnitude of any of the translation vectors in the 2D reciprocal lattice.
For this wavelength one of the allowed non-evanescent scattered wave directions is almost parallel to the plane of 2D layers.
The peak, observed in Fig.~\ref{kirch_dif_a}, is related to the smallest reciprocal translation vector with the length of $4\pi a/\sqrt{3}$.
This value corresponds to $\lambda=\sqrt{3}a/2$, which determines the left edge of the peak.
Peak amplitude is almost the same for different $a$ values due to the weak frequency dependence of the imaginary part of the tungsten dielectric function in the visible.

\begin{figure}
  \vspace{-10pt}
  \centering
  \scriptsize
  \includegraphics[width=1\linewidth]{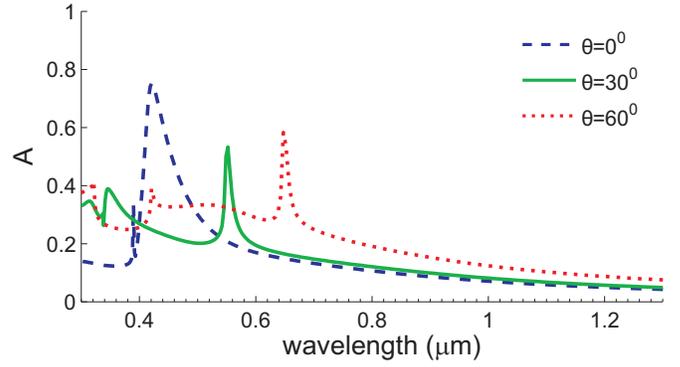}
  \caption{\small The absorption spectra of PC at different angles of incidence $\theta$ of an s-polarized wave. $a=0.45$ $\mu m$, $R=0.07$ $\mu m$ and $N=2$, $T=2400K$. The plane of incidence is determined by a normal to the layer and a 2D lattice primitive vector.}
  \label{kirch_dif_angle}
\end{figure}

Note that the critical wavelength value, corresponding to the diffraction edge, increases with the incident angle value.
As a result, the peak shifts towards the IR range while the angle of incidence increases (Fig.~\ref{kirch_dif_angle}).

\begin{figure}
  \vspace{-10pt}
  \centering
  \scriptsize
  \includegraphics[width=1\linewidth]{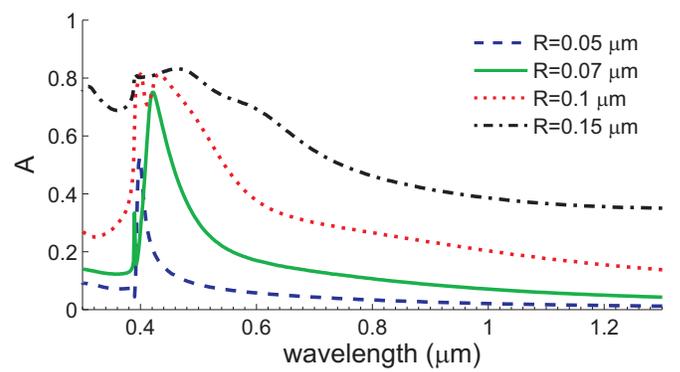}
  \caption{\small The absorption spectra of PC at normal incidence and varying sphere radius $R$. $a=0.45$ $\mu m$, $N=2$, $T=2400K$.}
  \label{kirch_dif_r}
\end{figure}

\begin{figure}
  \vspace{-10pt}
  \centering
  \scriptsize
  \includegraphics[width=1\linewidth]{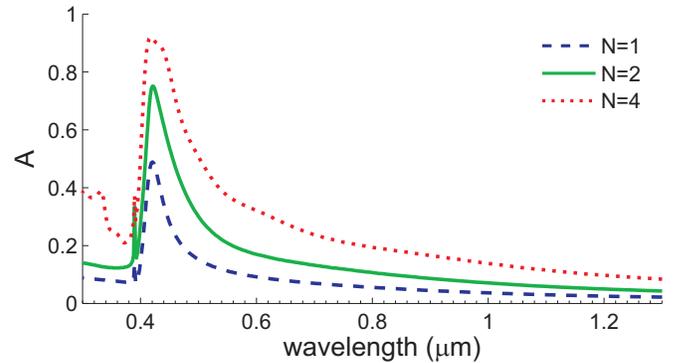}
  \caption{\small The absorption spectra of PC at normal incidence and varying number of layers $N$. $a=0.45$ $\mu m$, $R=0.07$ $\mu m$, $T=2400K$.
  }
  \label{kirch_dif_N}
\end{figure}

We calculate absorption spectra for different sphere radii $R$ and the number of layers $N$ (Figs.~\ref{kirch_dif_r},~\ref{kirch_dif_N}).
One can see that peak amplitude grows with the increase of $R$ and $N$, due to the larger amount of tungsten in the PC.
It leads to the higher energy emitted in the visible $J_{\rm vis}$ but lower efficiency $E = J_{\rm vis} / J_{\rm tot}$.

We now turn to the efficiency calculations for different values of the lattice constant $0.3\mu m\le a\le 0.7\mu m $, sphere radii $0.02\mu m \le R\le 0.2\mu m $, and the number of layers $1 \le N \le 4$.
The radiation flux is obtained by integrating  the absorption spectra over angles of incidence in the range $0^0 \le \theta < 90^0$, $0^0 \le \phi < 360^0$ according to~(\ref{pc_J}).
According to our calculations, for geometry with $a=0.45$ $\mu m$, $R=0.07$ $\mu m$, and $N=2$ (filling fraction $f \approx 2.2\%$) efficiency reaches the maximal value ($E \approx 15\%$) provided that $J_{\rm vis}$ is not less than $J_{\rm vis,W}/2$ ($J_{\rm vis} \sim 15 {\rm kWt/m}^2$).
As has been mentioned above, for the bare tungsten substrate $E_{\rm W}=5\%$, and $J_{\rm vis,W}=30 \mbox{kWt/m}^2$.
Therefore, the efficiency can be increased three times with $J_{\rm vis}$ lowered by a factor of 2 (due to using lower tungsten filling fraction), by using PC instead of the bulk tungsten.
\section{PC disorder}

\begin{figure}
  \vspace{-10pt}
  \centering
  \scriptsize
  \includegraphics[width=1\linewidth]{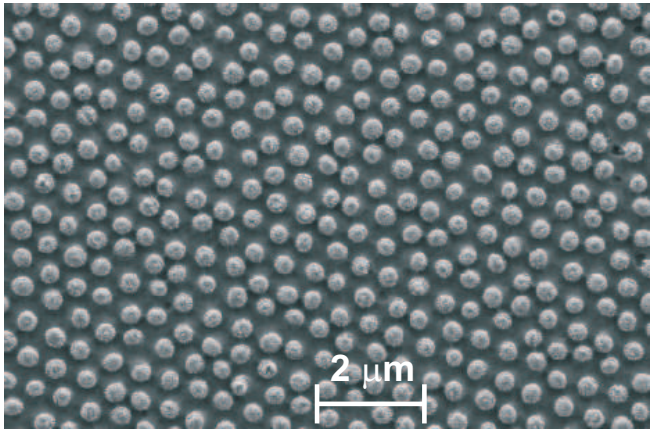}
  \caption{\small The top view of the PC sample with the triangular lattice period $a=0.55$ $\mu m$.}
  \label{kirch_exp_top}
\end{figure}

\begin{figure}
  \vspace{-10pt}
  \centering
  \scriptsize
  \includegraphics[width=1\linewidth]{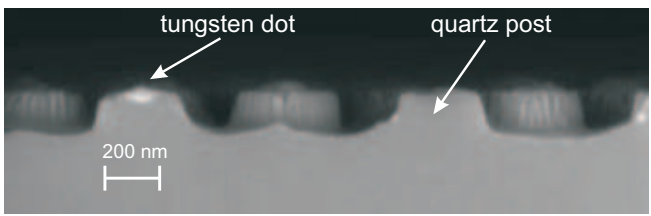}
  \caption{\small The side view of the PC sample with the triangular lattice period $a=0.55$ $\mu m$.}
  \label{kirch_exp_pr}
\end{figure}

In order to support the above results experimentally, a PC monolayer sample was fabricated on top of the quartz substrate. The sample consists of tungsten spherical segment caps on top of cylindrical quartz posts arranged in the triangular lattice with $a=0.55$ $\mu m$.
The cap base radius is $0.1$ $\mu m$, the cap height is $0.02$ $\mu m$, the quartz posts radius is $0.17$ $\mu m$ and the height is $0.15$ $\mu m$ (Figs.~\ref{kirch_exp_top},~\ref{kirch_exp_pr}).
We choose monolayer structure since it is relatively simple to fabricate, and the peak in the absorption spectra is present even if $N=1$ (Fig.~\ref{kirch_dif_N}).
The experiment has been carried out at the room temperature.

\begin{figure}
  \vspace{-10pt}
  \centering
  \scriptsize
  \includegraphics[width=1\linewidth]{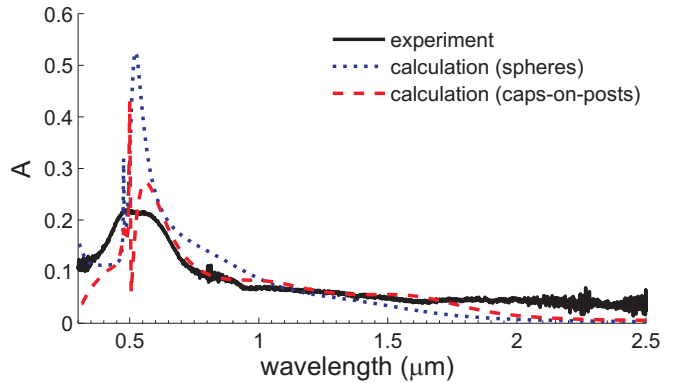}
  \caption{\small The tungsten PC monolayer absorption spectrum at normal incidence at the room temperature ($T=298K$). The PC parameters: $a=0.55$ $\mu m$, sphere segments radius $0.1$ $\mu m$, and height $0.02$ $\mu m$. The plotted curves correspond to: measurement; calculation for the perfect opal monolayer with $R=0.1$ $\mu m$; calculation for the PC with the caps-on-posts geometry of elements. The perfectly periodic lattice is assumed in calculations.}
  \label{kirch_exp_id}
\end{figure}

In Fig.~\ref{kirch_exp_id} we plot the comparison between the measured absorption of the fabricated PC sample and the calculated absorption of a) the PC monolayer consisting of tungsten spheres with $R=0.1$ $\mu m$, and b) the PC monolayer, consisting of tungsten caps on top of the quartz posts with dimensions corresponding to the experimental sample.
One can see that the peak, predicted by numerical calculations, is observed experimentally. In the case of the caps-on-posts PC the calculated absorption peak is smaller than in the case of spheres PC, due to the lower tungsten filling fraction in the former case. The even lower peak observed experimentally is caused by the lattice disorder and size dispersion of the PC elements.


\begin{figure}
  \vspace{-10pt}
  \centering
  \scriptsize
  \includegraphics[width=1\linewidth]{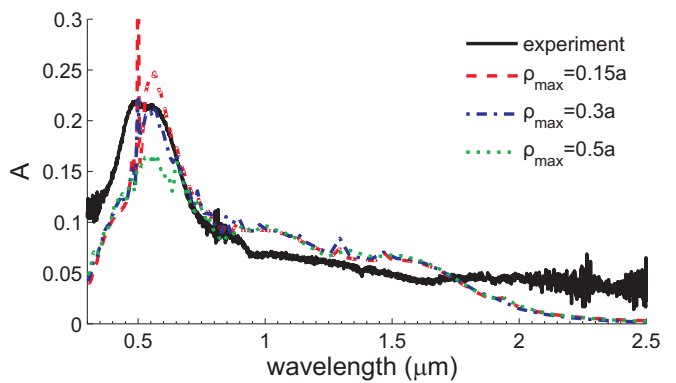}
  \caption{\small The tungsten PC monolayer absorption spectrum at normal incidence at the room temperature ($T=298K$). The PC parameters: $a=0.55$ ${\mu m}$, sphere segments radius $0.1$ $\mu m$, and height $0.02$ $\mu m$. Comparison of the measured spectrum to the calculation with both the size dispersion and the lattice disorder taken into account: results for the varying lattice disorder parameter $\rho_{\rm max}$ are presented (see text).
  }
  \label{kirch_exp_alpha}
\end{figure}

One can see from Fig.\ref{kirch_exp_top} that experimental sample is characterized by some level of deviation of position and size of PC elements from their "perfect" values. In order to reproduce the results for such type of "imperfect" PC numerically, we use the following scheme: a sufficiently large PC area is considered, with the XY dimensions much greater than the lattice constant. Each PC element is shifted randomly from its lattice node, the displacement being characterized by a vector with absolute value $\rho = \rho_{\rm max} \eta_{\rho}$ in the direction, specified by an angle $\phi = 2 \pi \eta_{\phi}$ with respect to the X axis, where $\rho_{\rm max}$ is the maximal possible displacement.
The volume of each PC element can vary around its mean value $V_0$ as $V = V_0+\Delta V_{\rm max}(2\eta_V-1)$, where $\Delta V_{\rm max}$ is the maximum possible variation, and $\eta_{\rho,\phi,V}$ are random numbers in $[0,1]$ interval.
According to this scheme, an imperfect PC is generated by using independent random values $\eta$ for assigning position and volume of each PC element according to the expressions given above.
Generated structure is placed into the FDTD computational cell with periodic boundary conditions in the X and Y directions.
This structure is randomly disordered at the small scale ($\sim a$), and uniform at the large scale ($\gg a$) that leads to the self-averaging of its optical properties~\cite{LAverage_eng}.
Absorption spectrum of this structure with appropriately chosen parameters $\rho_{\rm max}$ and $\Delta V_{\rm max}$ appears to be a good approximation for the absorption spectrum of the imperfect experimental PC.

In Fig.~\ref{kirch_exp_alpha} we present comparison of the calculated absorption spectrum for generated structure with $a=0.55\mu m$ and experimental data.
Computational cell lateral size is 8x8 PC unit cells with the period $a$ (further increase of lateral size does not change the results significantly).
The base radius of the sphere segments is varied, while its height is kept constant.
The maximum volume variation is set to $\Delta V_{\rm max}=0.5 V_0$.
The maximum shift $\rho_{\rm max} = 0.3 a$ gives the best agreement of the calculated peak height with the experimental data.
Note that the peak height decreases with the increase of parameter $\rho_{\rm max}$ corresponding to PC lattice disorder (Fig.~\ref{kirch_exp_alpha}).
At the same time, the IR part of the spectrum depends only on the size of the elements (e.g., see Fig.~\ref{kirch_dif_r}), since the wavelengths in IR are much greater than the lattice constant, and PC can be described as the effective medium in this spectral range.

In summary, the "imperfectness" of the experimental PC leads to decrease of the emission in the visible, however, the IR emission is still small.

\section{PC in a host matrix}

\begin{figure}
  \vspace{-10pt}
  \centering
  \scriptsize
  \includegraphics[width=1\linewidth]{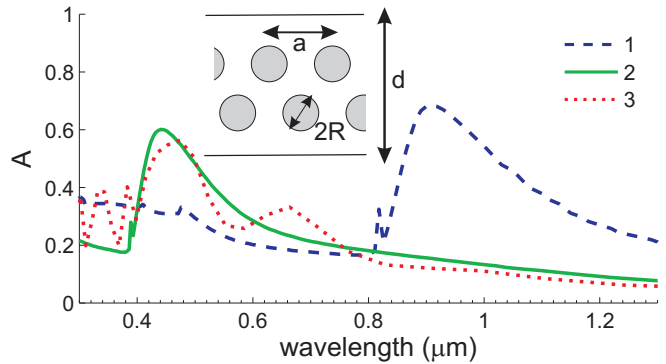}
  \caption{\small The absorption spectra of the two-layer PC at normal incidence ($T=2400K$). 1 - PC is embedded in hafnia at the room temperature, $a=0.45$ $\mu m$, $R=0.07$ $\mu m$; 2 - the same as in 1, with $a=0.214$ $\mu m$, $R=0.033$ $\mu m$; 3 - the finite thickness of the host is taken into account, $d$ is given by (\ref{eq:d}). Inset: Two-layer PC embedded in plane-parallel host with width $d$.}
  \label{kirch_host_d}
\end{figure}

The scattering elements of a PC should be embedded in a refractory material (host). There is only a small number of refractory materials, resistant to high temperatures and applicable for this purpose (i.~e., yttrium-stabilized zirconia and hafnia). However, when being heated, these materials emit in the IR. We find that it leads to a substantial efficiency reduction of the whole structure (PC and host). In this section we illustrate this result using an example of a tungsten two-layer PC embedded in the hafnia ($\rm HfO_2$) host (Fig.~\ref{kirch_host_d}, inset) with the width $d$:
\begin{equation}
\label{eq:d}
d = 2R+(1+1/\sqrt{3})a.
\end{equation}


We start with a non-absorbing host, taking the hafnia dielectric permittivity at room temperature ($\varepsilon=4.41$).
To avoid oscillations of the absorption curve caused by the finite thickness $d$ of the host slab, we consider the PC embedded in an infinite environment with $\varepsilon=4.41$.
The change of the environment dielectric function $\varepsilon$ from $1$ to $4.41$ leads to a red-shift of the absorption spectrum by a factor of $\sqrt{\varepsilon}$.
It is explained by the $\sqrt{\varepsilon}$-fold reduction of the wavelength in the external environment (compare blue dotted curves in Fig.\ref{kirch_dif_angle} and Fig.~\ref{kirch_host_d}, corresponding to the same PC geometry in vacuum and non-absorbing hafnia environment).
As a result, optimal parameters for the PC in host can be obtained by scaling down by $\sqrt{\varepsilon}$ that ones for vacuum environment (see green curve in Fig.~\ref{kirch_host_d}).

The finite thickness $d$ of the embedding host leads to the interference of the waves multiply reflected from the host-vacuum interfaces.
It leads to additional Fabry-Perot oscillations in the absorption spectrum.
However, it does not affect the location of the main absorption peak.
It can be seen in Fig.~\ref{kirch_host_d}, where the results for the PC in the infinite external environment (green curve) and in the finite slab of width $d$ defined by (\ref{eq:d}) (red dotted curve) are compared.

Hafnium becomes strongly absorbing at high temperatures. Its dielectric function at $T=2400K$ is well described by Lorentz approximation:
\begin{equation}
\label{hafnia_eps}
\varepsilon(\omega) = 1 + \frac{\Delta \varepsilon \omega_0^2}{\omega_0^2-2i\omega\gamma-\omega^2}.
\end{equation}
Here $\Delta \varepsilon = 3$, $\omega_0 = 1.2\cdot10^{16}\mbox{ rad/sec}$, $\gamma = 1.0\cdot10^{16}\mbox{ rad/sec}$.

\begin{figure}
  \vspace{-10pt}
  \centering
  \scriptsize
  \includegraphics[width=1\linewidth]{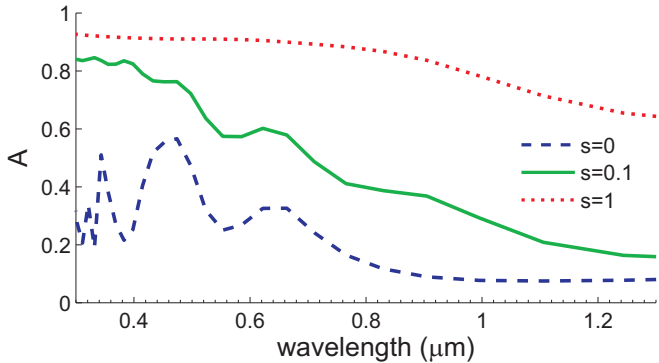}
  \caption{\small The absorption spectra of the PC at normal incidence ($T=2400K$). The PC parameters: $a=0.214$ $\mu m$, $R=0.033$ $\mu m$, $N=2$, $d$ is given by (\ref{eq:d}). Different values of the damping coefficient in the Lorentz formula for absorbing hafnia are considered (see text). The same FDTD parameters as on Fig.~\ref{kirch_dif_a} are used.}
  \label{kirch_host_abs}
\end{figure}

Effect of the host absorption can be examined by assuming $\gamma = s\cdot 10^{16}\mbox{ rad/sec}$ with parameter $s$ increasing gradually from 0 (zero absorption) to 1 (value from expression \ref{hafnia_eps}).
One can see from  Fig.~\ref{kirch_host_abs}, that for the small $s$ values the absorption spectrum is mainly determined by the tungsten spheres.
Increasing $s$ leads to the higher hafnia absorption in the IR and a significant decrease of the efficiency of the whole structure (7\% for $s=1$).
In summary, the photonic absorption peak is washed out by the absorption in the host matrix, necessary for embedding PC elements, so that a practical realization of PC-based light sources becomes difficult.

\section{Conclusions}

In this paper, we studied the possibility of using PCs as high-efficiency light sources. We considered different PC geometries (log-piles, direct and inverse opals) and showed that a direct opal geometry is more appropriate for our purpose. By changing geometrical parameters of a direct opal PC, one can achieve high selectivity of emission in the visible range. In real PC samples, a lattice disorder and a size dispersion of the constituent elements are present. This leads to a decrease of the peak emissivity in the visible, however, emissivity in the IR remains small.

For a practical application, the direct opal lattice needs to be embedded in a refractory host matrix. Known refractory materials are characterized by a high emissivity at the temperatures of interest. This leads to a lower PC peak emissivity in the visible. Therefore, implementing PC based light sources requires that low-absorbing refractory materials should be explored.

\section{Acknowledgments}

The work was partially supported by The Ministry of Science and Education of Russia under the contract No.16.523.11.3004.


\end{document}